\documentclass[11pt]{article}

\usepackage[utf8]{inputenc}
\usepackage[T1]{fontenc}
\usepackage{graphicx}
\usepackage{longtable}
\usepackage{wrapfig}
\usepackage{rotating}
\usepackage[normalem]{ulem}
\usepackage{amsmath}
\usepackage{amssymb}
\usepackage{capt-of}
\usepackage{hyperref}
\usepackage[citestyle=authoryear-icomp,bibstyle=authoryear,hyperref=true,backref=true,maxcitenames=3,url=true,backend=bibtex]{biblatex}
\usepackage{multirow}
\usepackage{amsmath}
\usepackage{morefloats}
\usepackage{setspace}
\usepackage{graphicx}
\usepackage{float}
\usepackage{changepage}
\author{Matt Brigida\thanks{SUNY Polytechnic Institute 100 Seymour Road, Utica, NY 13502, matthew.brigida@sunypoly.edu}}
\date{\today}
\title{Does Ecosystem Activity Validate TVL?}
\hypersetup{
 pdfauthor={Matt Brigida\thanks{SUNY Polytechnic Institute 100 Seymour Road, Utica, NY 13502, matthew.brigida@sunypoly.edu}},
 pdftitle={Does Ecosystem Activity Validate TVL?},
 pdfkeywords={},
 pdfsubject={},
 pdfcreator={},
 pdflang={English}}
\begin{document}

\maketitle
\begin{abstract}
Total value locked (TVL) is widely used to assess decentralized-finance ecosystems. Prior evidence finds that portfolios formed on TVL alone are spanned by the crypto market portfolio \parencite{brigida2025tvl}. This paper asks whether TVL becomes more informative when combined with measures of adoption, usage, development, protocol revenue, or social attention. I construct weekly value-weighted 2x3 portfolios that independently sort tokens on TVL-to-market-cap and one of seven ecosystem-activity ratios. Portfolio returns are computed from token prices using lagged-market-capitalization weights and evaluated against crypto market and size factors. Across the available 2022--2025 samples, none of the six on-chain double sorts rejects the benchmark after controlling for multiple testing. The same conclusion holds over 83 common weeks for which social-attention data are available. Several specifications reject at unadjusted levels, but none survives Holm adjustment. The social-attention rejection is only marginal under asymptotic inference and is insignificant under a moving-block bootstrap. Equal-weighted portfolios and formal parameter-stability tests lead to the same broad conclusion. The evidence therefore does not show that ecosystem activity reliably transforms TVL into an incremental return signal.
\end{abstract}
\vspace*{1cm}



\clearpage
\section{Introduction}
\label{sec:org08f5a1b}

Total value locked (TVL) is among the most visible measures of activity in decentralized finance. High TVL is commonly interpreted as evidence that users commit capital to an ecosystem, that its protocols are trusted, and that its associated token should benefit from adoption. Yet TVL is a stock of capital rather than a direct measure of tokenholder cash flow, and it is strongly related to ecosystem scale. Recent evidence accordingly finds that TVL alone does not produce positive market-adjusted returns \parencite{brigida2025tvl}.

This paper asks whether TVL contains more return information when it is supported by observable ecosystem activity. Two networks with the same TVL-to-market-cap ratio may differ substantially in active users, wallet creation, developer participation, decentralized-exchange volume, fees, transactions, or changes in social attention. These characteristics could distinguish productive capital from passively deposited capital. Conversely, if the information in these measures is already reflected in broad market and size exposures, conditioning on them should not produce abnormal returns.

Attention is a plausible conditioning variable. Evidence from equity markets links investor attention to buying behavior and returns \parencite{barber2008glitters}. Search intensity also provides a direct measure of investor attention \parencite{da2011search}. Research on cryptocurrency connects returns to investor attention \parencite{liu2021risks}. \textcite{urquhart2018attention} studies the sources of attention to Bitcoin. Other work links returns to media coverage \parencite{philippas2019media} and social-media activity \parencite{mai2018social}. Closest to this paper, \textcite{shen2019twitter} find that Twitter activity helps predict Bitcoin returns and volatility. I instead study the cross-section of tokens and treat attention as a characteristic that may interact with TVL.

I test this question with weekly independent 2x3 portfolio sorts. Tokens are first divided at the median of TVL relative to market capitalization. They are independently divided at the 30th and 70th percentiles of a second characteristic, also scaled by market capitalization. The intersections produce six value-weighted portfolios and two long-short legs. The first is a low-minus-high TVL-ratio leg. The second is a high-minus-low activity leg. I test whether a traded two-factor model containing the broad crypto market and a market-capitalization small-minus-big portfolio jointly spans these legs.

The analysis makes three contributions. First, it evaluates TVL jointly with seven measures spanning adoption, development, economic use, protocol revenue, and attention within a common portfolio framework. Second, it separates changes in token prices from changes in circulating supply by forming returns from prices and lagged market-capitalization weights. Third, it reports cross-sectional breadth and concentration, adjusts the family of spanning tests for multiple comparisons, and supplements heteroskedasticity-and-autocorrelation-consistent inference with a moving-block bootstrap and parameter-stability tests.

The principal result is negative but informative. In the full samples, the DEX-volume sort rejects the two-factor model at the unadjusted 5\% level, but no on-chain sort survives Holm adjustment across the six tests. Over the 83 common weeks beginning in September 2023, DEX volume and fees reject at unadjusted levels, while transactions and new X followers are marginal. None survives adjustment across seven tests. The New-X joint test has an asymptotic \(p\)-value of 0.052 and a block-bootstrap \(p\)-value of 0.106. Equal weighting changes which isolated specifications reject but does not produce a family-wise significant result. Formal stability tests likewise provide no adjusted evidence that the alphas changed during the attention-data period.

These findings qualify the idea that observable ecosystem growth validates TVL as a pricing signal. Some point estimates are economically large, but their statistical interpretation is sensitive to weighting, sample breadth, and the treatment of a family of related tests. The evidence supports neither an unconditional TVL interaction nor a distinct social-attention factor. Instead, it indicates that claims about on-chain fundamentals require careful return construction and inference before they can be interpreted as incremental expected-return signals.
\section{Data and Portfolio Construction}
\label{sec:org2302072}

The weekly sample runs from January 3, 2022 through July 28, 2025. The candidate universe contains tokens appearing among approximately the largest 100 cryptocurrencies during this period. Prices and market capitalizations come from the CoinMarketCap API. TVL, DEX volume, and fees come from DefiLlama. Active addresses, new wallets, active core developers, transactions, and new X followers come from Artemis.

The activity variables represent different dimensions of ecosystem strength. Active addresses and new wallets proxy for adoption. Active developers proxy for investment in protocol maintenance. DEX volume and transactions proxy for use. Fees proxy for protocol-level economic activity. Changes in X followers proxy for external attention. New-X data begin in August 2023. The pre-August values in the source file are an unavailable-data placeholder and are not treated as observations.

For each characteristic, I retain tokens with a positive TVL-to-market-cap ratio and finite lagged price, market capitalization, and characteristic data. Observed zero activity is retained rather than discarded and therefore enters the lower part of the activity distribution. Negative net changes in X followers are also retained. Bitcoin is excluded from the developer and DEX-volume sorts because those measures do not have the same economic interpretation for Bitcoin as for smart-contract ecosystems. It is retained in all other sorts.\footnote{Bitcoin is excluded only from the developer and DEX-volume sorts. When it is retained, the developer conclusion is unchanged. The DEX sort rejects at unadjusted levels over 78 of the 83 common-sample weeks. Its moving-block-bootstrap \(p\)-value over those 78 weeks is 0.050, and the result remains one of a family of related specifications.}
\subsection{Independent 2x3 Sorts}
\label{sec:orgf20e005}

For token \(i\) at the end of week \(t-1\), define

$$V_{i,t-1}=\frac{TVL_{i,t-1}}{MCAP_{i,t-1}}, \qquad
A^F_{i,t-1}=\frac{F_{i,t-1}}{MCAP_{i,t-1}},$$

where \(F\) is one of the seven activity characteristics. Tokens are independently assigned to a smaller or larger \(V\) group using the weekly median and to low, medium, or high \(A^F\) groups using the 30th and 70th percentiles. This produces portfolios \(SL\), \(SM\), \(SH\), \(BL\), \(BM\), and \(BH\), where \(S\) and \(B\) refer to smaller and larger TVL-to-market-cap ratios rather than token market capitalization.

Portfolio returns are calculated from token price returns rather than changes in aggregate market capitalization.

$$R_{P,t}=\sum_{i\in P_{t-1}}w_{i,t-1}
\left(\frac{Price_{i,t}}{Price_{i,t-1}}-1\right), \qquad
w_{i,t-1}=\frac{MCAP_{i,t-1}}{\sum_{j\in P_{t-1}}MCAP_{j,t-1}}.$$

This definition prevents token issuance, burns, and changes in circulating supply from being mechanically counted as investment returns. It also ensures that all formation information and portfolio weights are lagged.

The first test leg averages the three smaller-\(V\) portfolios and subtracts the three larger-\(V\) portfolios,

$$LH^{TVL}_t=\frac{R_{SL,t}+R_{SM,t}+R_{SH,t}}{3}
-\frac{R_{BL,t}+R_{BM,t}+R_{BH,t}}{3}.$$

The second is a high-minus-low activity leg conditional on the TVL ratio,

$$HML^F_t=\frac{R_{SH,t}+R_{BH,t}}{2}
-\frac{R_{SL,t}+R_{BL,t}}{2}.$$

Table 1 reports the available weeks, mean returns, and construction diagnostics. The median eligible universe ranges from 16 tokens for new wallets to 54 for developers. The New-X universe has a median of 26. Because the design creates six portfolios, some values remain small. The final row reports, for each week, the largest constituent weight in any of the six values and then takes its time-series median. These values show that at least one cell is usually concentrated, motivating equal-weighted robustness tests.

\begin{table}[htbp]
\caption{Descriptive statistics and construction diagnostics. Returns are weekly percentages. Universe is the median number of eligible tokens per week. Min. cell is the median across weeks of the smallest of the six cell counts. Max. weight is the time-series median of the largest constituent weight in any cell. Add. denotes active addresses. Devs denotes active developers. DEX denotes decentralized-exchange volume.}
\centering
\begin{tabular}{lrrrrrrr}
\hline
 & Add. & Devs & DEX & Wall. & Fees & Trans. & New X\\
\hline
\(N\) & 187 & 187 & 169 & 163 & 186 & 187 & 97\\
Mean \(LH^{TVL}\) & 0.59 & 0.43 & 0.92 & -0.08 & 0.76 & 0.80 & 0.39\\
Mean \(HML^F\) & -0.12 & -0.26 & 0.96 & 0.65 & 0.34 & 0.63 & 1.57\\
Median universe & 40 & 54 & 46 & 16 & 42 & 38 & 26\\
Median min. cell & 3 & 4 & 1 & 2 & 3 & 3 & 2\\
Median max. wt. & 97.1\% & 79.2\% & 100\% & 96.4\% & 95.7\% & 95.9\% & 98.5\%\\
\hline
\end{tabular}
\end{table}
\section{Empirical Method}
\label{sec:org4e5372e}

The benchmark contains two traded factors. \(MKT_t\) is the lagged-market-cap-weighted return on the broad token universe less the weekly U.S. Treasury-bill return. \(SIZE_t\) is the value-weighted return on the smallest market-cap quartile minus that on the largest quartile. This parsimonious specification follows evidence that market exposure is central to cryptocurrency returns \parencite{liu2021risks}. It also follows evidence that common risk factors explain important parts of the cryptocurrency cross-section \parencite{liu2022common}. All returns, including the risk-free rate, are expressed in percentage points.

For each characteristic \(F\), I estimate the system

$$r^P_t=\alpha^P+\beta^P_{MKT}MKT_t+\beta^P_{SIZE}SIZE_t+u^P_t,
\qquad P\in\{LH^{TVL},HML^F\}.$$

The spanning null sets \(\alpha^{LH^{TVL}}=\alpha^{HML^F}=0\). I estimate the equations by OLS and compute a joint Wald statistic using a Bartlett heteroskedasticity-and-autocorrelation-consistent covariance estimator with bandwidth four \parencite{newey1987simple}. The statistic has an asymptotic \(\chi^2(2)\) reference distribution. This is a HAC Wald test of zero alphas. It is not the classical finite-sample GRS test \parencite{gibbons1989test}.

The tables report each alpha with its HAC standard error, the joint statistic, and its \(p\)-value. Because the sorts share observations and the TVL leg, I treat them as a family of related hypotheses and report Holm-adjusted \(p\)-values \parencite{holm1979simple}. Holm adjustment controls the family-wise error rate without requiring independent tests. For the shorter matched sample, I additionally report a 499-replication moving-block residual bootstrap with four-week blocks \parencite{kunsch1989jackknife}. The resampling is circular \parencite{politis1992circular}.

The New-X data permit 97 usable weeks on their own. To compare all seven sorts on identical dates, I use the intersection of their available observations and the benchmark factors. This produces 83 common weeks from September 4, 2023 through July 28, 2025. Matching dates ensures that differences across columns are not driven by different return weeks. It does not equalize the number of tokens within each sort.
\section{Results}
\label{sec:org11a5e51}

\subsection{Full Samples}
\label{sec:org71b9351}

Table 2 reports results for the six on-chain characteristics over their available samples. DEX volume rejects the joint spanning null at the unadjusted 5\% level (\(p=0.048\)). Its Holm-adjusted \(p\)-value is 0.291. No other joint test is significant at 5\%, and none of the six survives the family-wise adjustment. Transactions produces an individually significant activity-leg alpha of 0.73\% per week, but its joint \(p\)-value is 0.107. The full-sample evidence therefore does not establish that any on-chain activity ratio reliably adds return information to TVL-to-market-cap.

\begin{table}[htbp]
\caption{Full-sample traded-factor results. Alphas and HAC standard errors (in parentheses) are weekly percentages. \(J\) tests the joint null that both alphas are zero. Holm \(p\) adjusts the six joint tests for family-wise multiple comparisons. \(R^2\) is pooled across the two equations.}
\centering
\begin{tabular}{lrrrrrr}
\hline
 & Add. & Devs & DEX & Wall. & Fees & Trans.\\
\hline
\(\alpha_{LH^{TVL}}\) & 0.383 & 0.353 & 0.798 & -0.109 & 0.516 & 0.485\\
 & (0.349) & (0.433) & (0.393) & (0.384) & (0.310) & (0.380)\\
\(\alpha_{HML^F}\) & 0.109 & -0.057 & 0.883 & 0.525 & 0.251 & 0.732\\
 & (0.297) & (0.504) & (0.475) & (0.352) & (0.346) & (0.356)\\
\(J\) & 1.253 & 0.785 & 6.055 & 2.883 & 2.771 & 4.468\\
\(p\) & 0.534 & 0.675 & 0.048 & 0.237 & 0.250 & 0.107\\
Holm \(p\) & 1.000 & 1.000 & 0.291 & 0.946 & 0.946 & 0.536\\
\(R^2\) & 0.106 & 0.125 & 0.023 & 0.047 & 0.092 & 0.112\\
\(N\) & 187 & 187 & 169 & 163 & 186 & 187\\
\hline
\end{tabular}
\end{table}
\subsection{Common Attention-Data Sample}
\label{sec:org13b45ae}

Table 3 estimates every sort on the same 83 weeks. At unadjusted asymptotic levels, DEX volume (\(p=0.044\)) and fees (\(p=0.012\)) reject the spanning null. Transactions and New X followers are marginal, both with \(p=0.052\). After Holm adjustment, the smallest \(p\)-value is 0.086 for fees. None rejects at 5\%.

The New-X activity-leg alpha is economically large at 1.77\% per week, but its HAC standard error is 0.75\%. The two-alpha joint test is marginal rather than conventionally significant, and its block-bootstrap \(p\)-value is 0.106. Thus the data do not identify a social-attention interaction that is robust to small-sample inference and multiple testing. More generally, the columns with low unadjusted \(p\)-values are not stable across weighting schemes, as shown below.

\begin{table}[htbp]
\caption{Common-sample results. Every sort uses the same 83 weeks from 9/4/2023 through 7/28/2025. Alphas and HAC standard errors (in parentheses) are weekly percentages. Boot. \(p\) is the four-week moving-block-bootstrap \(p\)-value. Holm \(p\) adjusts the seven asymptotic joint tests.}
\centering
\begin{tabular}{lrrrrrrr}
\hline
 & Add. & Devs & DEX & Wall. & Fees & Trans. & New X\\
\hline
\(\alpha_{LH^{TVL}}\) & 0.689 & -0.158 & 1.049 & -0.087 & -0.350 & 0.515 & -0.180\\
 & (0.552) & (0.418) & (0.524) & (0.480) & (0.414) & (0.685) & (0.432)\\
\(\alpha_{HML^F}\) & 0.523 & -0.609 & 1.695 & 0.718 & 0.688 & 1.068 & 1.766\\
 & (0.439) & (0.489) & (0.731) & (0.563) & (0.503) & (0.440) & (0.751)\\
\(J\) & 1.918 & 2.573 & 6.266 & 2.069 & 8.794 & 5.913 & 5.912\\
\(p\) & 0.383 & 0.276 & 0.044 & 0.355 & 0.012 & 0.052 & 0.052\\
Boot. \(p\) & 0.456 & 0.372 & 0.118 & 0.366 & 0.038 & 0.126 & 0.106\\
Holm \(p\) & 0.829 & 0.829 & 0.262 & 0.829 & 0.086 & 0.262 & 0.262\\
\(R^2\) & 0.151 & 0.269 & 0.030 & 0.063 & 0.154 & 0.154 & 0.130\\
\hline
\end{tabular}
\end{table}
\subsection{Robustness and Limitations}
\label{sec:org26c603b}

Table 4 summarizes weighting and inference robustness. Under equal weighting, transactions and DEX volume reject in the full sample at unadjusted levels. DEX volume also rejects in the common sample at an unadjusted level. Their Holm-adjusted values are 0.086, 0.233, and 0.051, respectively. The equal-weight common-sample DEX rejection also weakens to 0.050 in the block bootstrap. The identity of the lowest-\(p\) sort therefore changes with portfolio weighting, and no specification provides clear family-wise evidence at the 5\% level.

\begin{table}[htbp]
\caption{Robustness of joint-test \(p\)-values. VW and EW denote value- and equal-weighted test portfolios. Common denotes the 83-week attention-data sample. Boot. denotes the moving-block bootstrap. Values are unadjusted. No family shown in the table rejects at 5\% after Holm adjustment.}
\centering
\begin{tabular}{lrrrrrr}
\hline
 & Full VW & Full EW & Common VW & Common Boot. & Common EW & EW Boot.\\
\hline
Add. & 0.534 & 0.293 & 0.383 & 0.456 & 0.539 & 0.660\\
Devs & 0.675 & 0.389 & 0.276 & 0.372 & 0.973 & 0.976\\
DEX & 0.048 & 0.047 & 0.044 & 0.118 & 0.007 & 0.050\\
Wall. & 0.237 & 0.301 & 0.355 & 0.366 & 0.143 & 0.198\\
Fees & 0.250 & 0.289 & 0.012 & 0.038 & 0.248 & 0.344\\
Trans. & 0.107 & 0.014 & 0.052 & 0.126 & 0.368 & 0.456\\
New X & -- & -- & 0.052 & 0.106 & 0.329 & 0.484\\
\hline
\end{tabular}
\end{table}

I also test whether the two alphas change after the beginning of the common attention-data sample while allowing both factor loadings to change. Fees has an unadjusted alpha-stability \(p\)-value of 0.031, but its Holm-adjusted value across the six on-chain sorts is 0.188. The other unadjusted values range from 0.219 to 0.937. The evidence therefore does not establish a structural change in pricing relationships.

Retaining Bitcoin in the developer sort leaves the full and common-sample conclusions unchanged. Retaining it in the DEX sort produces an unadjusted \(p\)-value of 0.015 in the full sample and 0.015 over 78 of the 83 common-sample weeks. The block-bootstrap \(p\)-value over those 78 weeks is 0.050. This sensitivity is consistent with Bitcoin's dominant weight and atypical DEX activity and reinforces the need to interpret isolated DEX rejections cautiously.

Several limitations remain. First, the weekly sample is short relative to traditional asset-pricing panels, especially for social attention. Second, the value-weighted cells are concentrated. The largest constituent in at least one cell often receives nearly all of that cell's weight. Equal weighting mitigates but does not eliminate the small number of tokens in some cells. Third, the benchmark is intentionally parsimonious. Momentum, liquidity, and other crypto factors could further explain the isolated alphas. Fourth, activity and attention measures may contain reporting error and can be affected by incentives, marketing, exchange listings, or automated behavior. Finally, the analysis identifies return associations rather than causal effects of ecosystem activity.
\section{Conclusion}
\label{sec:org3650bd7}

This paper tests whether measures of ecosystem activity transform TVL-to-market-cap into a return signal not captured by broad crypto market and size factors. Across six on-chain characteristics, none of the full-sample joint spanning tests survives adjustment for multiple comparisons. The same conclusion holds when all seven sorts are estimated over the common 83-week attention-data sample.

The strongest unadjusted results are not uniform across specifications. DEX volume is nominally significant in several tests, fees rejects in the value-weighted common sample, and transactions rejects under full-sample equal weighting. New X followers produces a large activity-leg alpha, but its joint rejection is marginal under asymptotic inference, insignificant under the block bootstrap, and insignificant after family-wise adjustment. Equal-weighted portfolios and formal stability tests do not yield a consistently priced interaction.

The evidence therefore does not support treating TVL interactions as reliable expected-return factors. This conclusion does not imply that TVL or ecosystem activity lacks economic information. Rather, within this sample and portfolio framework, the information does not consistently appear as abnormal returns relative to a simple traded-factor benchmark. Longer attention histories, broader token coverage, and richer benchmark models are necessary to determine whether any of the nominal DEX, fee, transaction, or attention estimates persist out of sample.

\clearpage

\printbibliography
\end{document}